\documentclass{article}

 \usepackage[nonatbib, final]{nips_2018}

\usepackage[utf8]{inputenc} 
\usepackage[T1]{fontenc}    
\usepackage{hyperref}       
\usepackage{url}            
\usepackage{enumerate}
\usepackage{amsmath}
\usepackage{dcolumn}
\usepackage{booktabs}       
\usepackage{amsfonts}       
\usepackage{nicefrac}       
\usepackage{microtype}      
\usepackage{float}

\title{A Second Order Cumulant Spectrum Test That a Stochastic Process is Strictly Stationary and a Step Toward a Test for Graph Signal Strict Stationarity}

\author{
Denisa ~Roberts\thanks{Corresponding author is Denisa Roberts at d.roberts@vt.edu, Machine Learning Scientist at Amazon at the time of the submission, currently AI Researcher unaffiliated with Amazon.} \\
  Amazon \\
  New York, NY 10001 \\
  \texttt{rdenisa@amazon.com} \\
   \And
   Douglas ~Patterson \\
  Department of Finance\\
  Virginia Tech\\
  Blacksburg, VA 24061 \\
  \texttt{amex@vt.edu} \\
}

\begin{document}

\maketitle

\begin{abstract}

This article develops a statistical test for the null hypothesis of strict stationarity of a discrete time stochastic process in the frequency domain. When the null hypothesis is true, the second order cumulant spectrum is zero at all the discrete Fourier frequency pairs in the principal domain. The test uses a window averaged sample estimate of the second order cumulant spectrum to build a test statistic with an asymptotic complex standard normal distribution. We derive the test statistic, study the properties of the test and demonstrate its application using 137Cs gamma ray decay data. Future areas of research include testing for strict stationarity of graph signals, with applications in learning convolutional neural networks on graphs, denoising, and inpainting.

\end{abstract}

\section{Introduction and related work}
\label{sec:intro}

A stochastic process $\{X(t)\}$, $t=0..T$, is called strictly stationary when the entire family of its finite dimensional distribution presents invariance under a common translation in the time structure \cite{brillinger2001time}. The concept of stationarity of a time series plays a significant role in time series analysis and its applications, and is an assumption for building time series forecasting models and drawing the correct inferences (for example financial returns prediction, demand forecasting, human activity prediction). One can obtain meaningful sample statistics of the time series (mean, variances, correlations with other variables) that can be useful descriptors of future behavior only if the time series is approximately stationary. There are two paradigms for analyzing time series: the time domain and the frequency domain. In the time domain, one considers the observed data directly and typically makes conjectures about its moments. In the frequency domain, one decomposes the time series into underlying frequencies and makes conjectures about spectra and cumulant spectra. This paper operates in the frequency domain and presents a novel statistical time series method based on the second order cumulant spectrum to test the hypothesis that an observed discrete time series $\{x(t_n)\}$, $t_n=n\tau$, with sampling rate $\tau$, is strictly stationary. 

The classical idea of time series stationarity has been extended to the graph signal processing (GSP) domain and we aim to extend our test to GSP as well. One potential application is for learning convolutional neural networks (CNN) in high-dimensional irregular domains such as social networks, brain connectomes, log data on telecommunication networks, gene data on biological regulatory networks, or word embeddings represented by graphs as in \cite{defferrard2016convolutional}. The intention is to generalize CNNs ability to learn local, stationary and compositional features from low-dimensional regular grids, where image, video and speech are represented, to irregular structures represented by graphs \cite{defferrard2016convolutional}. The major bottleneck of generalizing CNNs to graphs is the definition of localized graph filters which can be learnt \cite{defferrard2016convolutional}. Graph signal processing domain can lend mathematical tools such as spectral graph theory to handle the challenge. Certain local properties such as smoothness and local stationarity are tied to effective learning on graphs. The domain of graph signal processing generally provides tools for modeling data defined on irregular grids \cite{ortega2018graph}. As another example of graph stationarity application, \cite{perraudin2017stationary} leverage the idea of stationary graph signals to derive Wiener-type estimation procedures of noisy and partially observed signals that can be used for denoising and inpainting. They show how stationarity improves over classical graph models and Gaussian MAP estimators \cite{perraudin2017stationary}.

We develop a test for strict stationarity based on the second order cumulant spectrum

\begin{equation}
\label{eq:1}
K\left(f_1, f_2\right) = \sum_{n_1 = -\infty}^{\infty}\sum_{n_1 = -\infty}^{\infty}E\left[x\left(t_{n_1}\right)x\left(t_{n_2}\right)\right]exp\left[-2\pi\left(f_1t_{n_1} + f_2t_{n_2}\right)\right],
\end{equation}
defined for the Fourier frequencies $\{- f_0 \le f_1 \le f_0, - f_0 \le f_2 \le f_0 \}$,  where $f_0 = 1 / \left(2 \tau\right)$. 

For simplification we will refer to the test employing the cumulant spectrum as Cum2 for the remainder of this article. The Cum2 test is different from tests for structural change of a linear time series model, such as CUSUM charts (\cite{andreou2009structural} gives a review). The Cum2 test is able to detect changes due to several types of violations of the null hypothesis of strict stationarity. Tests of a unit root with or without drift against an alternative hypothesis of trend stationarity, can be found, for example, in \cite{dickey1979distribution} and \cite{phillips1988testing}. In \cite{dwivedi2011test} the authors derived a test of second order stationarity based on the correlation of discrete Fourier transforms. A time series is called second-order stationary, weakly stationary or stationary in the wide-sense \cite{brillinger2001time} if $E\{x\left(t_m\right)\} $ is constant and its autocovariance $c_x\left(t_m\right)$ depends only on the difference $\left(t_{n+m} - t_{n}\right)$, $c_x\left(t_m\right) = c_x\left(t_{n+m} - t_{n}\right)$. A strictly stationary series with finite second order moments is second-order stationary \cite{brillinger2001time}. Hence our test is more general than the test proposed in \cite{dwivedi2011test}. Furthermore, our test has different power from Hinich-Wild \cite{hinich2001testing} test of strict stationarity using the signal coherence method. The Hinich-Wild test has power against the alternative of a randomly modulated periodic process. Our test has more general power and is not limited to such an alternative. Cum2 is also different from the spectral correlation detection in test in \cite{baugh1994detection}. Since the space of non strictly stationary processes is vast it is difficult to specify the full range of alternatives.

We present the test development in Section \ref{sec:cum2}, evaluate the test empirical size and power on synthetic data in Section \ref{sec:empirical}, demonstrate its application in Section \ref{sec:application} and conclude in Section \ref{sec:conc}.

\section{The Cum2 test for strict stationarity}
\label{sec:cum2}

In Equation \ref{eq:1} the cumulant spectrum $K$ is defined for frequency values in the square $\{- f_0 \le f_1 \le f_0, - f_0 \le f_2 \le f_0 \}$,  where $f_0 = 1 / \left(2 \tau\right)$ is the Nyquist frequency. Because of Fourier frequencies symmetries \cite{bloomfield2004fourier}, we focus the study of the second order cumulant spectrum for the set of fundamental frequencies in the principal domain only, namely the triangle $\{0 < f_1 \le f_0, -f_1 < f_2 \le f_1\}$. From \cite{hinich1994higher}, if the time series $\{x\left(t\right)\}$ is strictly stationary then the second order cumulant spectrum for a pair of frequencies in the principal domain

\begin{equation} \label{eq:4}
K\left(f_1, f_2 \right)= E[X(f_1)X(f_2)] = 
\left\{
	\begin{array}{ll}
		2S\left( f _1\right) + O\left( 1\right) & \mbox{if } f_1+f_2 = \mbox{ 0 mod  1 }\\			
		O\left(1\right) & \mbox{ otherwise, }
	\end{array}
\right.
\end{equation}

\noindent where $X(f_i)$ denote the discrete Fourier transform. In Equation \ref{eq:4}, $K\left(f_1, -f_1\right) = \delta\left(f_1\right)S\left(f_1\right)$,
where $\delta\left(f_1\right)$ is the Dirac delta function, and $S\left(f_1\right) = \sum_{m = -\infty}^{\infty}c_x\left(t_m\right)\exp\left(-i2\pi f_1t_m\right)$ is the spectrum at frequency $f_1$ and  $c_x\left(t_m\right)$ is the autocovariance of the process. In Equation \ref{eq:4}, $X(f_i)$ are the discrete Fourier transforms at fundamental frequencies in the principal domain. In summary, if a time series is strictly stationary, then its second order cumulant spectrum $K\left(f_1, f_2\right) = 0$  for all frequency pairs $\left(f_1, f_2 \right)$ in the principal domain \cite{hinich1994higher}.

We now use the properties of the second order cumulant spectrum to construct the test for the null hypothesis that a stochastic process $\{x(t_n)\}$ is strictly stationary. We assume that the time series has been demeaned, detrended, prewhitened and trimmed, as is typical in frequency domain applications. The test uses frame averaged cumulant spectra at fundamental frequency pairs and has an asymptotic complex normal distribution. To get frame averaged estimates for the cumulant spectrum $K(f_1, f_2)$ and for the spectrum $S(f)$, where $(f_1, f_2)$ are frequencies in the principal domain, we use the windowing trick: partition the time series into frames of equal length, calculate the cumulant spectrum and the spectrum for each frame and then get averages of spectral quantities of interest. A time series of length $T$ is partitioned into $P = [T/ L]$ complete non overlapping frames of length $L$. We omit the last frame if it has less than $L$ observations. Then the frame averaged cumulant spectrum estimate is $\hat{K}\left(f_1 , f_2 \right) = \frac{1}{P} \sum_{p=1}^{P} K_p\left(f_1 , f_2 \right).$

For computational simplicity, we express the fundamental frequencies as $f = k/L$, with $0 < k_1 \le L/2$, $-k_1 \le k_2 \le k_1$.  We similarly estimate the window averaged spectrum, $\hat{S}\left(f_k\right)$. From the theory in \cite{hinich1994higher} and \cite{brillinger2001time}, if $L$ and $P$ are sufficiently large, the expected value of the frame averaged spectrum is equal to its theoretical value up to $O(L^{-1})$, $E\left[\hat{S}\left(f_k\right)\right] = S\left(f_k\right) + O\left( L^{-1}\right)$. Similarly, under strict stationarity, from \cite{hinich1994higher}, $E\left[\hat{K}\left(k_1 , k_2 \right)\right] = K\left(k_1 , k_2 \right) + O\left(L^{-1}\right).$ Furthermore, the variance of $\hat{K}$, considering $P$ frames, is equal to $P^{-1}S\left({k_1}\right)S\left({k_2}\right)$ as $\left(L, P \to \infty\right)$ from \cite{hinich1994higher}. We define the normalized second order cumulant spectrum as $\hat{\Gamma} \left(k_1, k_2 \right) = \frac{\hat{K}\left(k_1 , k_2 \right)}{\sqrt{S\left(f_{k_1}\right)S\left(f_{k_2}\right)}}.$ This normalization is similar to the normalization of the bispectrum to yield the skewness function. We make a further simplifying assumption used in \cite{hinich1998frequency} that the time series has been prewhitened, hence the theoretical spectrum $S$ is assumed constant across frequencies and, without loss of generality, equal to one. With this assumption, the standard error of $\hat{K}$ becomes $1/\sqrt P$. Then the estimate for $\Gamma$, $\hat{\Gamma}$ simplifies to $\sqrt{P}\hat{K}$. We now construct the complex valued Cum2 test statistic quantity $Y$ as $Y\left(k_1, k_2\right) = \sqrt {2P} \left[ \hat{\Gamma}\left(k_1, k_2 \right) - \Gamma \left(k_1, k_2 \right) \right]$. Assuming $P \to \infty$, from the central limit theorem, the real and imaginary parts of  $Y\left(k_1, k_2\right)$ are asymptotically independent Gaussian variates with zero means and unit variances as $\left(L, P \to \infty\right)$ if the process is strictly stationary. The Kolmogorov-Smirnov (KS) one sample test is a straightforward way to test the null hypothesis that the random variables $\Re{(Y(k_1,k_2))}$ and $\Im{(Y(k_1,k_2))}$ follow a standard normal distribution. Let $F_{emp}\left(x\right)$ denote the empirical cumulative distribution of $\Re{(Y)}$, and $\Im{(Y)}$ respectively. We calculate  $F_{emp}\left(x\right)$  as the fraction of values that are less than a given value $\emph{x}$ in the unit interval. In our case the vector $\mathbf x$ represent the real and imaginary parts for the Cum2 test statistic and the cumulative distribution of interest is that of a standard normal variable. 

We are working to extend the Cum2 theory to graph signal processing. In \cite{girault2017towards}, Girault et. al introduce the concept of local estimate of the graph power spectrum and its use as an estimate for the global spectrum. We draw the analogy to the window averaged estimate of the spectrum used as a building block in Cum2 test. In \cite{girault2017towards} a window on the graph is defined as a L-hop neighborhood, with a span $L$. Let's consider $P$ graph windows of span $L$ and define for Laplacian eigenvalues $k_1$ and $k_2$ the window averaged second order cumulant spectrum as $E[GFTX_{k1}*GFTX_{k2}]$ where $GFTX_k$ are the graph Fourier transform (GFT) components under the isometric translation operator from \cite{girault2015stationary} for eigenvalues $k$. Then a GFT Cum2 test statistic may be defined as a normalized graph window averaged cumulant spectrum. The formal definitions of graph higher order cumulants, test statistic, windowing technique, test properties and asymptotic distribution are still to be formulated but we believe this to be a step toward a formal test statistic for graph stationarity, which, to our best knowledge, does not exist just yet.

\section{Empirical evaluation of the Cum2 test size and power}
\label{sec:empirical}

Next we evaluate the size and power of the test on synthetic data as the percentage samples out of 10,000 generated samples where the test rejects stationarity. We expect to see empirical test size close to the ground truth test size and large power. An R package \cite{R15} to implement Cum2 and calculate a p value is available in supplementary materials. We first generate strictly stationary data (white noise) in order to evaluate the empirical size of Cum2. At 5 percent ground truth size of test, the empirical size of test for different sample sizes $T$ are 0.062 for $T=250$, 0.063 for $T=1000$ and 0.059 for $T=5000$. Next we evaluate the empirical power of the test. There are many alternatives of non stationarity that can be specified for Cum2. We restrict our attention to alternatives of unit root and varying second moment. We first generate data $x(t) = \lambda  y(t) + u(t)$ to include a unit root component $y(t)=\sum_{j=1}^{t}v(j)$ and a white noise stationary component $u(t)$. The test rejects the null of strict stationarity with power of 0.080 for $T=250$, 0.420 for $T=1000$ and 0.997 for $T=5000$ respectively. We then generate synthetic data with a varying second moment, $x(t) =  \sigma_t u(t)$, where $u(t)$ is white noise and $\sigma_t^2$ is the variance. The variance is one under the null of stationarity and follows a piecewise linear trend under alternative, such that $\sigma^2_t = {\sigma_0}^2 + \left( {\sigma_1}^2 - {\sigma_0}^2\right)\left(t-m\right)  \left(1-m\right)^{-1}$ for $t \ge m$ and ${\sigma_0}^2$ for $t < m$, with $\sigma_0 = 1$ and $\sigma_1 = d\sigma_0$. Table \ref{tab:tabseven} illustrates the power of the test to detect piecewise linear trend in second moment (increase if $d = 4$ and decrease if $d = 0.25$) that starts at different points in the sample (half way if $m = 0.5$ and late if $m = 0.9$), for different sample sizes, $T = \{250, 1000, 5000\}$, and at five percent ground truth size of the test. 
    
 \begin{table}[!htbp] \centering 
 \caption{Empirical power of test to detect piecewise linear trend in the second moment} 
 \label{tab:tabseven} 

\begin{tabular}{@{\extracolsep{0pt}} D{.}{.}{-3} D{.}{.}{-3} D{.}{.}{-3} D{.}{.}{-3} D{.}{.}{-3} D{.}{.}{-3} D{.}{.}{-3} D{.}{.}{-3} D{.}{.}{-3} D{.}{.}{-3} } 
\toprule
\multicolumn{1}{c}{m} & \multicolumn{1}{c}{d} & \multicolumn{1}{c}{Cum2.250} & \multicolumn{1}{c}{Cum2.1000} & \multicolumn{1}{c}{Cum2.5000} \\ 
\midrule
\multicolumn{1}{c}{0.5} & \multicolumn{1}{c}{4} & 0.273  & 1  & 1  \\ 

\multicolumn{1}{c}{0.5} & \multicolumn{1}{c}{0.25} & 0.078  & 0.378  & 0.956 \\ 

 \multicolumn{1}{c}{0.9} & \multicolumn{1}{c}{4} & 0.447  & 1  & 1  \\ 

\multicolumn{1}{c}{0.9} & \multicolumn{1}{c}{0.25} & 0.066  & 0.084 & 0.105 \\ 
 
\bottomrule

\end{tabular} 
\end{table}

\section{Gamma ray decay application}
\label{sec:application}

We applied the test to a dataset of measurements of 137Cs gamma ray decay. The data were collected using a 137Cs source and a NaI detector in a Pb shielding enclosure to minimize the background contribution. The NaI detector produces a light pulse which is proportional to the amount of energy that the gamma ray deposits in the detector. For each light pulse the total light output is measured using a photomultiplier and a multichannel analyzer which bins the outputs into discrete bins depending on the light output. The time between consecutive events was computed, resulting in a series of 999,508 inter-arrival times. The laws of physics predict that the pulse sequence is a stationary Poisson process and thus the inter-arrival time sequence is a stationary random process. The Cum2 test p-value is 0.26 for a frame length of $L = 200$, failing to reject the null hypothesis that the process is strictly stationary, as expected.

\section{Conclusion}
\label{sec:conc}

In this article we used the frame averaged estimates of the second order cumulant spectrum calculated for pairs of Fourier frequencies in the principal domain to build a statistical test of the null hypothesis that a stochastic process is strictly stationary. We based the test development on the property that the second order cumulant spectrum of a time series in the frequency domain is zero under strict stationarity. We studied the empirical size and power of the test and demonstrated its performance in an application to gamma ray decay data. Future paths of research include the extension of the Cum2 theory to develop a test for strict stationarity of a graph signal with applications in learning CNNs on graphs, denoising, inpainting, weather prediction, and social networks. We relate to the theory developed in \cite{girault2015stationary}. In the graph signal processing domain, we define a graph signal $X$ measured at a sequence of vertices in $V$, where $G=(V,E)$ is a a weighted undirected graph. Girault {\cite{girault2015stationary} introduces an isometric graph translation operator $T_{G}=\exp(-it\sqrt{La})$ where $La$ is the graph Laplacian matrix. Then a stochastic graph signal is strictly stationary if $X$ is equivalent in distribution to $T^{t}_{G}X$ for all $t$. The graph Fourier transform $GFTX=X(\exp{\left(-it\sqrt{\lambda}\right)})$ is then defined as a projection of X on the eigenvectors of the Laplacian ${La}$. Girault \cite{girault2015stationary} introduces the concept of graph wide sense stationarity, equivalent to classical signal processing wide sense (second order) stationarity, if $E[GFTX_l]=0$ for $\lambda_l \neq 0$ and $E[GFTX_{k_1}*GFTX^{*}_{k_2}] = 0$ for $\lambda_{k_1} \neq \lambda_{k_2}$. No formal test statistic is defined. We propose extending Cum2 test to graph stationarity testing.

\bibliographystyle{unsrt}
\bibliography{my_bib}

\section*{Supplementary materials}

\subsection*{Sensitivity analysis}

We include a sensitivity analysis of the Cum2 test to see how the sample size and frame length $L$ affect the empirical test size. We generated white noise and evaluated Cum2 empirical size under different sample size and frame length scenarios. Simulation results in Table \ref{tab:tabthree} show that it is best to have sample sizes above 500, at least 50 frames and a frame length of at least 10. A rule of thumb is to set $L$ and $P$ equal to the square root of sample size $T$. 

\begin{table}[H] \centering 
  \caption{Empirical Test Size for Various Sample Sizes T and Frame Lengths L (corresponding number of frames P=[T/L]). Nominal Test Size of 0.05.} 
  \label{tab:tabthree} 
\scriptsize
\begin{tabular}{@{\extracolsep{0pt}} D{.}{.}{-3} D{.}{.}{-3} D{.}{.}{-3} D{.}{.}{-3} D{.}{.}{-3} } 
\\[-1.8ex]\hline \\[-1.8ex] 
\multicolumn{1}{c}{L} & \multicolumn{1}{c}{T.250} & \multicolumn{1}{c}{T.500} & \multicolumn{1}{c}{T.1000} & \multicolumn{1}{c}{T.5000} \\ 
\hline \\[-1.8ex] 
\multicolumn{1}{c}{10} & 0.062 & 0.059 & 0.062 & 0.058 \\ 
\multicolumn{1}{c}{20} & 0.062 & 0.058 & 0.060 & 0.058 \\ 
\multicolumn{1}{c}{50} & 0.195 & 0.084 & 0.063 & 0.057 \\ 
\multicolumn{1}{c}{100} & 1 & 0.716 & 0.183 & 0.059 \\ 
\multicolumn{1}{c}{200} & 1 & 1 & 1 & 0.122 \\ 
\hline \\[-1.8ex] 
\end{tabular} 
\end{table} 

\subsection*{Code}

An R package implementing the test can be installed from \url{https://github.com/D-Roberts/statiotest}.

\end{document}